\documentstyle[12pt,a4wide]{article}
\input epsf
\newcommand{\be}{\begin{equation}}
\newcommand{\ee}{\end{equation}}

\renewcommand{\epsilon}{\varepsilon}
\font\mybb=msbm10 at 11pt

\def\bb#1{\hbox{\mybb#1}}

\def\bR {\bb{R}}

\renewcommand{\theequation}{\arabic{section}.\arabic{equation}}
\newcommand{\news}{\setcounter{equation}{0}}
\def\ben{\begin{equation}}
\def\een{\end{equation}}
\def\bea{\begin{eqnarray}}
\def\eea{\end{eqnarray}}
\begin{document}

\title{\vskip -0cm
\bf \Large \bf Schr\"odinger-Chern-Simons Vortex Dynamics\\[30pt]
\author{Steffen Krusch and Paul Sutcliffe\\[10pt]
\\{\normalsize{\sl Institute of Mathematics,
University of Kent, Canterbury, CT2 7NF, U.K.}}\\
{\normalsize{\sl Email : S.Krusch@kent.ac.uk}}\\
{\normalsize{\sl Email : P.M.Sutcliffe@kent.ac.uk}}\\[0.5cm]}}
\date{April 2006}
\maketitle

\begin{abstract}
We study the motion of vortices in the planar Ginzburg-Landau model
with Schr\"odinger-Chern-Simons dynamics. We compare the moduli space 
approximation with the results of numerical simulations of the full field
theory and find that there is agreement if the coupling constant
is very close to the critical value separating Type I from Type II
superconductors. However, there are significant qualitative differences
even for modest deviations from the critically coupled regime. 
Radiation effects produce forces which are of the same order of magnitude 
as the intervortex force and therefore have a significant impact on 
vortex motion. We conclude that the moduli space approximation does not
provide a good description of the dynamics in this regime. 
\end{abstract}

\newpage

\section{Introduction}\news
The planar Ginzburg-Landau model provides a good mathematical
description of the static properties of 
vortices in thin superconductors, but in order to
describe vortex motion this needs to be supplemented by some
appropriate dynamics. There are several possibilities for the form
of this dynamics and it is currently difficult to determine which
of these is the closest to modelling real superconductors, since 
in experimental situations the vortices are often pinned, so
it is difficult to identify the underlying dynamical behaviour.

It has been argued \cite{Ai}
that at very low temperatures dissipation
can be ignored and vortex motion is orthogonal to the force acting,
so that two vortices circulate around each other, as in a fluid.
An interesting Lagrangian field theory
to describe this type of behaviour has been proposed
by Manton \cite{Ma10} in which the potential part is the usual 
Ginzburg-Landau energy and the kinetic terms are linear in the
time derivatives of the complex scalar field and contain a Chern-Simons
term. The equation of motion for the complex scalar field 
is a gauged Schr\"odinger equation, hence the name 
Schr\"odinger-Chern-Simons dynamics. It is Manton's model that we
shall study in this paper.

At critical coupling, which separates Type I from Type II
superconductors, minimizers of the Ginzburg-Landau potential energy
are static solutions of the Schr\"odinger-Chern-Simons equations,
so there is no vortex motion. Thus to investigate vortex dynamics
the model needs to be studied away from critical coupling.

The moduli space approximation \cite{Ma5} to soliton dynamics, 
in which the field theory dynamics is approximated by motion on a 
finite dimensional manifold, has been applied to 
Schr\"odinger-Chern-Simons vortex dynamics \cite{Ma10,SpR}.
The moduli space used is the manifold of static vortices at critical
coupling, so it should be most accurate when the coupling constant
is close to its critical value, though it can not equal this value
if there is to be vortex motion, as mentioned above.
In particular, it predicts that indeed two vortices circulate around
each other at constant speed, and provides an expression for the
period of this motion in terms of quantities defined on the moduli
space. Polygonal configurations of $N$ vortices can also be studied and 
the moduli space approximation predicts that 
for well separated vortices these symmetric configurations
are stable if and only if $N<6.$

In this paper we perform numerical simulations of the full field theory 
dynamics and
compare the results with the predictions of the moduli space approximation.
For two vortices we find that there is a good 
agreement when the coupling constant
is very close to the critical value. However, there are significant 
qualitative differences even for modest deviations away from critical
 coupling. We find that two vortices move on spirals, rather than
circles, and spiral in or out depending upon the value of the
coupling constant. This behaviour is due to radiation effects 
which are neglected in the moduli space approximation but turn
out to produce forces which are of the same order of magnitude as the 
intervortex force. Polygonal arrangements also exhibit the same
spiral phenomenon and the stability properties do not always agree
with the moduli space predictions.   
 
\section{The Schr\"odinger-Chern-Simons Model}\news\label{sec-model}
The Ginzburg-Landau potential energy is given by
\begin{equation}
\label{VGL}
V = \frac{1}{2} \int \left(B^2 + \overline{D_i \phi} D_i \phi
+ \frac{\lambda}{4}(1 - \overline{\phi} \phi)^2 \right) d^2 x,
\end{equation}
where $\phi$ is a complex scalar field, representing the electron
pair condensate, and 
the covariant derivative is
$D_i \phi = \partial_i \phi - i a_i \phi,$  formed using the 
abelian gauge potential $a_i$
with an associated magnetic field 
\begin{equation}
B = \partial_1 a_2 - \partial_2 a_1.
\end{equation}
In (\ref{VGL}), and the following, the summation convention applies
to the two spatial indices only, ie. $i=1,2.$

The finite energy topological solitons of this model are known as vortices.
The vortex number $N$ is a topological degree and is equal to the 
total magnetic flux in units of $2 \pi$,
\begin{equation}
N = \frac{1}{2 \pi} \int B~ d^2 x.
\end{equation} 
$N$ is also equal to the total number of zeros of the complex scalar
field $\phi$ counted with multiplicity. 
These zeros can be interpreted as the vortex positions when the vortices  
are well separated.

The positive parameter $\lambda$ plays a crucial role in the properties
of the model and its vortex solutions.
For $\lambda<1$ the superconductor is of Type I and 
the potential energy $V$ of two vortices is an increasing 
function of their separation ie. vortices attract.

For $\lambda > 1$ the superconductor is of Type II
and the potential energy of two vortices
is a decreasing function of their separation ie. vortices repel. 
At the critical coupling $\lambda=1,$ separating Type I from Type II
superconductors, the potential energy is independent of
their separation. Moreover, at critical coupling,
and for all positive vortex numbers $N,$ the parameter space 
of all minimal energy $N$-vortices forms a
$2N$-dimensional smooth manifold, known as the moduli space $M_N$. 
In this case the potential energy of any $N$-vortex solution is equal to
$\pi N$ for arbitrary values of the vortex positions.

At critical coupling the second order field equations
which follow from the variation of the potential (\ref{VGL}) 
can be reduced to two first order Bogomolny equations 
\begin{eqnarray}
D_1 \phi + i D_2 \phi &=& 0, \label{Bog1}\\
B - \frac{1}{2} (1 - \overline{\phi} \phi) &=& 0, \label{Bog2}
\end{eqnarray} 
and these will play an important role later.

The Schr\"odinger-Chern-Simons model
introduced by Manton \cite{Ma10} is a non-dissipative model for
vortex motion in thin superconductors. The model is defined by
the Lagrangian
\be
\nonumber
L= \int \left( \frac{i}{2} \left( \overline{\phi} D_0
\phi - \phi \overline{D_0 \phi} \right) 
+ (B a_0 + e_1 a_2 - e_2 a_1)  - a_0 \right) d^2 x -V,
\label{LSCS}
\ee
where we have scaled some possible parameters to convenient values and
$e_i$ is the electric field  $e_i = \partial_0 a_i -\partial_i a_0.$ 
Note that the term $B a_0 + e_1 a_2 - e_2 a_1$ is simply the
Chern-Simons term written out explicitly and that the $-a_0$
term is required to allow the possibility of a condensate at infinity.
The potential term $V$ of the Lagrangian (\ref{LSCS}) is the
Ginzburg-Landau energy (\ref{VGL}). This Lagrangian is both gauge invariant and
Galilean invariant, when one takes into account a constant external transport
 current. Although this is an important feature of the model 
we shall not be concerned with this aspect here.

The equations of motions which follow from (\ref{LSCS}) are given by
\bea
\label{Schr}
i  D_0 \phi &=& -\frac{1}{2} D_i D_i \phi
- \frac{\lambda}{4} (1 - \overline{\phi} \phi) \phi,\\
\label{Amp}
-\epsilon_{ij} \partial_j B &=&  
\frac{i}{2} \left(\overline{\phi} D_i \phi - \phi \overline{D_i \phi}
\right) + 2 \epsilon_{ij} e_j,\\
\label{Gauss}
 B &=& \frac{1}{2}(1 - \overline{\phi} \phi).
\eea
The first
equation is a gauge covariant nonlinear Schr\"odinger equation.
The second equation is an Amp\`ere equation,
where the total current is the sum of the usual supercurrent 
and a Hall current orthogonal to the electric
field. The third equation is a generalized Gauss law,
which contains no time derivatives. It can be interpreted as a
constraint on the initial data since it can be shown that 
the first two equations already imply that it is conserved, namely,
\begin{equation}
\frac{\partial}{\partial t} \left(
B - \frac{1}{2} (1 - \overline{\phi} \phi) \right) = 0.
\end{equation}
Therefore, once equation (\ref{Gauss}) is satisfied for the
initial data, equations (\ref{Schr}) and (\ref{Amp}) 
guarantee that it is satisfied for all later times. 

As the Lagrangian is linear in time derivatives then the kinetic energy
makes no contribution to the conserved energy, which is simply the 
potential energy $V,$ whose conservation is easily checked using the
equations of motion.

If the Schr\"odinger equation (\ref{Schr}) and the Amp\`ere equation
(\ref{Amp}) are considered for static fields, with a vanishing
time component for the gauge potential $a_0=0,$ then these two equations
reduce to the two second order field equations obtained from the
variation of the Ginzburg-Landau energy (\ref{VGL}). However, 
away from critical coupling, these static Ginzburg-Landau vortices
will not satisfy the Gauss law (\ref{Gauss}) and hence are not static
solutions of the Schr\"odinger-Chern-Simons equations. The exception is at 
critical coupling $\lambda=1$ where Ginzburg-Landau vortices also
satisfy the Bogomolny equations (\ref{Bog1}) and (\ref{Bog2}), the
second of which is precisely the Gauss law (\ref{Gauss}).   

The first issue to consider is therefore the static solutions of
the Schr\"odinger-Chern-Simons model away from critical coupling.
We shall restrict to axially symmetric solutions, since it is expected 
that all static solutions will have axial symmetry when $\lambda\ne 1.$
We work in the radial gauge $a_r = 0,$ with $a_\theta(r)$ and $a_0(r)$ 
functions of the radius $r$ only, and the complex scalar field having
the standard $N$-vortex form 
$ \phi = f(r) {\rm e}^{i N \theta},$ where $f(r)$ is the radial profile
function. 

Substituting this axial ansatz into the field equations (\ref{Schr}),
(\ref{Amp}) and (\ref{Gauss}) yields the following set of ordinary
differential equations
\begin{eqnarray}
\label{aSchr}
& &f^{\prime \prime} +\frac{1}{r} f^\prime 
- \frac{1}{r^2} \left( N-a_\theta \right)^2 f + 
\frac{\lambda}{2} \left(1-f^2 \right) f + 2 a_0 f
= 0, \\
\label{aAmp}
& &\left( \frac{1}{r} a_\theta^\prime \right)^\prime 
+ \frac{1}{r} \left(N-a_\theta \right) f^2 - 2 a_0^\prime 
= 0,\\
\label{aGauss}
& & \frac{2}{r} a_\theta^\prime + f^2-1 = 0,
\end{eqnarray}
where $\prime$ denotes differentiation with respect to $r$. The 
boundary conditions for the fields are 
$f(0) = 0$, $f(\infty) = 1$, $a_\theta(0) = 0$, $a_\theta(\infty) = N$
and $a_0'(0) = 0.$ Note that from (\ref{aSchr}) these boundary
conditions automatically imply that $a_0(\infty)=0.$ 

\begin{figure}[!ht]
\begin{center}
\leavevmode \vskip -1cm \epsfxsize=15cm\epsffile{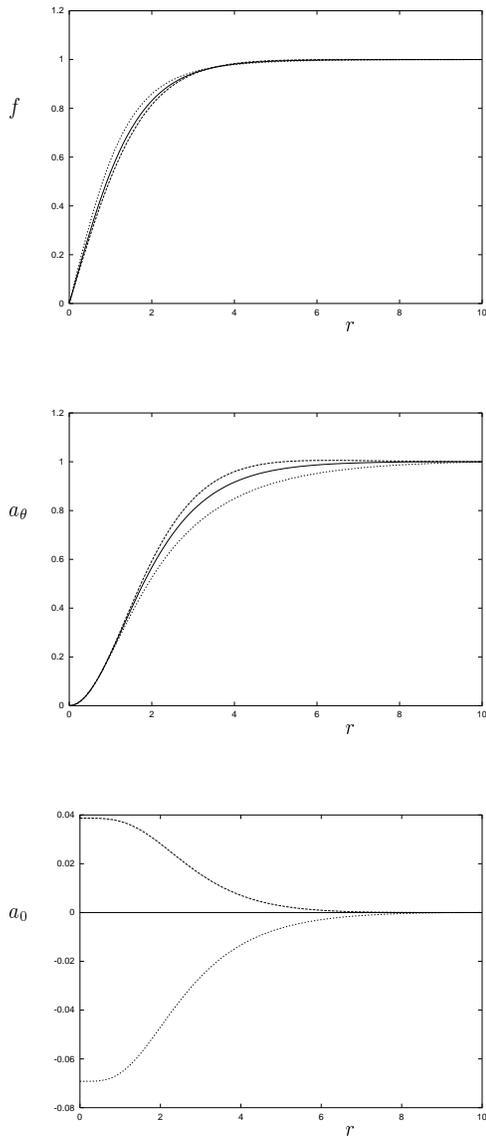} \vskip -6cm
\caption{Graphs of the functions $f(r)$, $a_\theta(r)$ and $a_0(r)$ for
couplings  $\lambda=1$ (solid curves), 
  $\lambda=0.5$ (dashed curves),
$\lambda=2$ (dotted curves).
\label{fig-radial}}
\end{center}
\end{figure}

These equations are solved using a gradient flow method with a 
fictitious energy constructed from the square of the field equations.
The results for the $N=1$ vortex are displayed in Fig.~\ref{fig-radial},
for three values of the coupling constant; $\lambda=1$ (solid curves), 
  $\lambda=0.5$ (dashed curves), $\lambda=2$ (dotted curves).
The critical coupling case $\lambda=1$ is shown
for comparison, where 
$a_0=0$ and the functions $f(r)$ and $a_\theta(r)$ are the usual fields of the
 Ginzburg-Landau vortex. 
The qualitative features of both the profile function $f(r)$ and the 
angular component of the gauge field $a_\theta(r)$ do not vary significantly
away from critical coupling. If $\lambda<1$ then $f(r)$
is slightly wider and $a_\theta(r)$ is a little narrower, with the
opposite being true for $\lambda>1,$ that is, $f(r)$ is narrower 
and $a_\theta(r)$ is wider. The most significant feature is the
behaviour of the temporal component of the gauge potential $a_0,$ which
is positive for $\lambda<1$ and negative for $\lambda>1.$
Thus, away from critical coupling, vortices not only have a magnetic field 
but also a tiny electric field. For a single vortex the electic field is radial
and positive if $\lambda<1,$ whereas it is negative if $\lambda>1.$

\begin{figure}
\begin{center}
\leavevmode \vskip -3cm \epsfxsize=12cm\epsffile{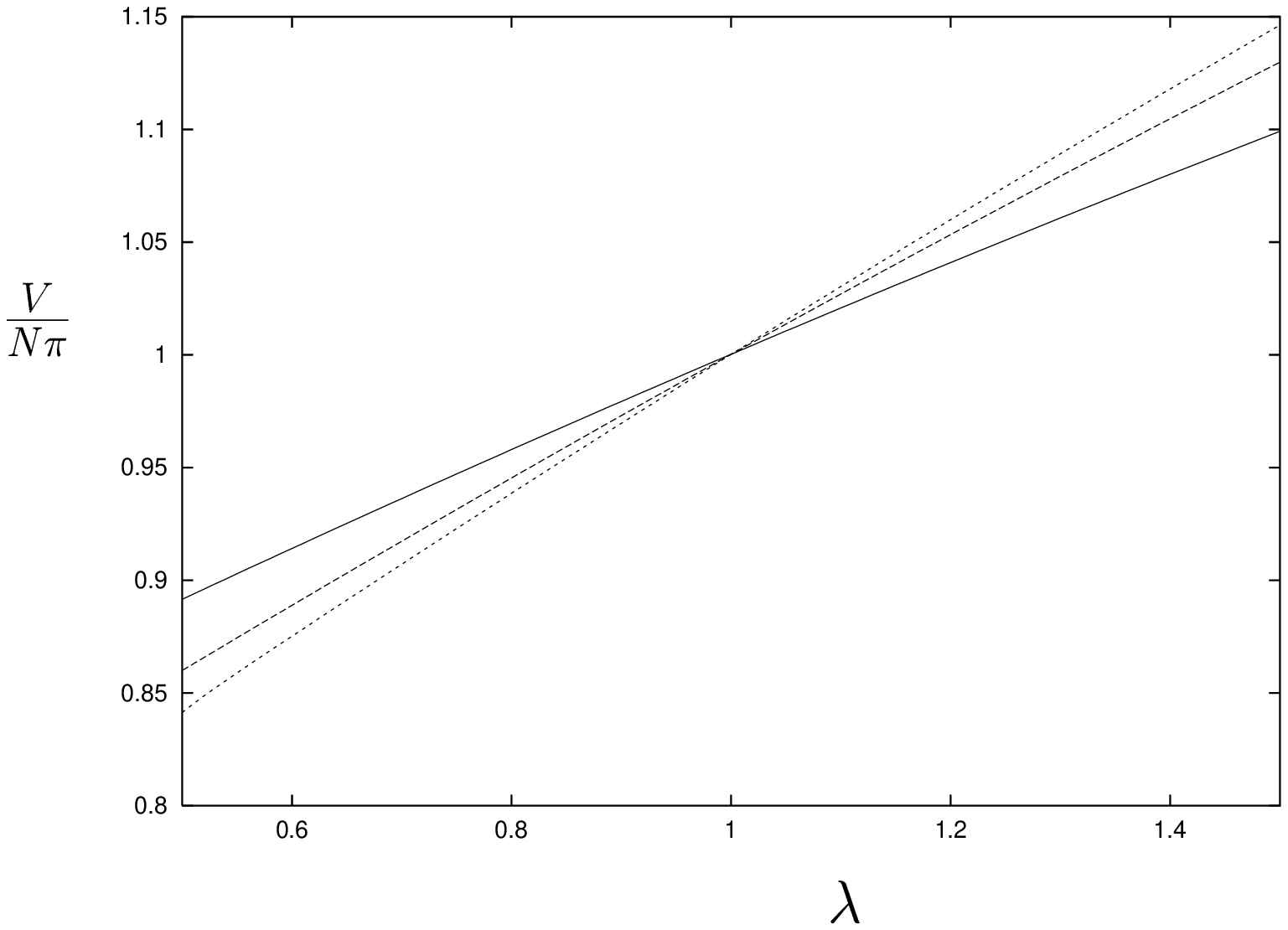} \vskip -8cm
\caption{The energy per vortex (in units of $\pi$)
 as a function of the coupling $\lambda$ for the
axial $N$-vortex with $N=1$ (solid curve), $N=2$ (dased curve), 
$N=3$ (dotted curve).}
\label{fig-energy}
\end{center}
\end{figure}

As we have mentioned earlier, the conserved energy in the 
Schr\"odinger-Chern-Simons model is the Ginzburg-Landau energy $V.$
However, for $\lambda\ne 1$ the static fields are not those of the
Ginzburg-Landau model, so the energy dependence on $\lambda,$ for
varying vortex numbers $N,$ can not be inferred from knowledge
of the Ginzburg-Landau model. In Fig.~\ref{fig-energy} we plot 
the energy per vortex (in units of $\pi$) of the axial $N$-vortex 
as a function of $\lambda$ for
$N=1$ (solid curve), $N=2$ (dashed curve), 
$N=3$ (dotted curve). We have also computed similar results for
 larger values of $N,$ but for clarity they are not shown on this plot.

Despite the above comment we find that
the qualitative behaviour of the energy is similar to
Ginzburg-Landau vortices. In particular, for multi-vortices we see that 
for $\lambda < 1$  the energy
of $N$ well separated vortices is higher than the energy of an axially
symmetric charge $N$ vortex, so vortices are attractive in this sense.
For $\lambda > 1$ vortices are repulsive, in that the energy of
$N$ well separated vortices is lower than the charge
$N$ axial configuration. At critical coupling, $\lambda=1,$ the static 
vortices coincide with Ginzburg-Landau vortices, so the energy per
vortex equals $\pi$ for all $N,$ as it does for all vortices in
the $2N$-dimensional moduli space $M_N$.

The above features of the vortex interaction energy will play an 
important role in the dynamics of vortices, as we shall see later.

\section{The Moduli Space Approximation}\news

The main idea of the moduli space approximation \cite{Ma5} is to project the
dynamics of the full field theory onto a suitable finite dimensional space 
of field configurations. In the simplest situation, such as vortices
at critical coupling or BPS monopoles, this finite dimensional space
is the moduli space of static minimal energy solutions, with a given
soliton number. This approach is well established 
(for a review see \cite{book}) for 
relativistic Lagrangians, where the dynamics is second order in time.
Restricting the field theory Lagrangian to the moduli space produces 
(upto an irrelevant constant)
a purely kinetic Lagrangian on the moduli space, 
which may be interpreted as the moduli space metric.
The slow motion of solitons is approximated by geodesic motion
with respect to this metric. For relativistic dynamics of 
Ginzburg-Landau vortices in the abelian Higgs model 
the validity of this approximation has been proved with mathematical
rigour \cite{Stu2}.

 A slightly more complicated situation arises
if a theory is considered at parameter values for which the
moduli space of static solutions is not large enough to describe
solitons with arbitrary positions, for example, vortices in the
abelian Higgs model away from critical coupling. In such situations
the dynamics can be truncated to motion on the moduli space of the
theory at critical coupling, but now there will also be a potential
energy function on the moduli space, so the dynamics is no longer
described by geodesic motion. This approach has been applied in
different models \cite{Su1,Shah}, and there is a good agreement with
results obtained from numerical simulations of the full 
relativistic field theory dynamics. For relativistic vortex dynamics
near to critical coupling rigourous error estimates can again be
obtained \cite{Stu2}.    

Schr\"odinger-Chern-Simons vortex dynamics is first order in time,
so initial conditions consist only of vortex positions.
Furthermore, at critical coupling the Ginzburg-Landau vortices
are static solutions, so there is no vortex motion. Thus, in
constrast to relativistic dynamics, it is 
necessary to move away from critical coupling in order for there
to be any vortex dynamics to study. Manton has argued \cite{Ma10}
that it is possible to use moduli space techniques to
describe Schr\"odinger-Chern-Simons vortex dynamics
close to critical coupling, and this has been further investigated 
in some detail by Rom\~ao and Speight\cite{SpR}.
 We summarise the results of the relevant calculations below.

Let $q^\alpha$, for $\alpha=1,\ldots,2N,$ be real coordinates on the
moduli space $M_N,$ for example the real and imaginary parts of the 
vortex positions when space is identified with the complex plane. 
As the  Schr\"odinger-Chern-Simons Lagrangian (\ref{LSCS})
depends only linearly on time derivatives then the Lagrangian restricted
to the moduli space must have the form
\be 
L_{\mbox {red}}={\cal A}_\alpha(q)\dot q^\alpha - V_{\mbox{red}}(q),
\label{redlag}
\ee  
where ${\cal A}_\alpha(q)$ may be interpreted as the components of
a $U(1)$ connection form ${\cal A}={\cal A}_\alpha dq^\alpha$ on
$M_N.$ The first order equations of motion which follow from (\ref{redlag})
are 
\be
{\cal F}_{\alpha\beta}\dot q^\alpha=
-\frac{\ \partial V_{\mbox{red}}}{\partial q^\beta},
\label{msdyn}
\ee
where ${\cal F}=d{\cal A}$ is the curvature form of ${\cal A}.$
In order to study the details of this reduced dynamics we first
need a better understanding of the moduli space $M_N$.

By considering the Bogomolny equations (\ref{Bog1}) and (\ref{Bog2}) 
Taubes \cite{Tau1} showed that 
it is possible to eliminate the gauge potential as follows. Define
\begin{equation}
h = \log |\phi|^2,
\label{htophi}
\end{equation}
which is gauge invariant and finite, except at the zeros of
$\phi$. Using the first Bogomolny equation (\ref{Bog1}), 
it can be shown that the
magnetic field can be expressed as
\begin{equation}
B = - \frac{1}{2} \nabla^2 h.
\end{equation}
The second Bogomolny equation then becomes
\begin{equation}
\nabla^2 h + 1 - {\rm e}^h = 0.
\label{heqn2}
\end{equation}
Recall $h$ has logarithmic singularities at the zeros of $\phi$. This
can be accounted for by including delta-function sources into the
equation,
\begin{equation}
\nabla^2 h + 1 - {\rm e}^h = 4 \pi \sum\limits_{r=1}^N 
\delta(z - Z_r),
\label{heqn}
\end{equation}
where $\delta$ is the two dimensional delta-function and 
$Z_r,$ with $r=1,\ldots,N.,$ 
are the complex positions of the zeros of $\phi$ in $\bR ^2,$ which is
identified with the complex plane. It can be shown that $h$ has the
following expansion around a zero $z=Z_r$
\begin{equation}
h(z,\overline{z}) = 2 \log |z - Z_r| + a_r +
\frac{1}{2} \overline{b}_r (z-Z_r) + 
\frac{1}{2} b_r(\overline{z} - \overline{Z}_r) + O(|z-Z_r|^2).
\label{hexpan}
\end{equation}
It turns out that the coefficients $b_r$ contain all the necessary
information to reconstruct the geometry of the moduli space. 

Following the work of Strachan \cite{Str}, Samols \cite{Sam} 
was able to show that the moduli space metric derived from
vortex dynamics in the relativistic abelian Higgs model
is given by
\begin{equation}
ds^2 = \pi \sum\limits_{r,s=1}^{N} \left(\delta_{rs} 
+ 2 \frac{\partial b_s}{\partial Z_r}\right) dZ_r d\overline{Z}_s,
\end{equation}
which is a K\"ahler metric. The corresponding K\"ahler form is
\begin{equation}
\omega = \frac{i \pi}{2} 
\sum\limits_{r,s=1}^{N} 
\left( \delta_{rs} + 2 \frac{\partial b_s}{\partial Z_r} \right)
dZ_r \wedge d \overline{Z}_s.
\end{equation}
and it turns out that the curvature form ${\cal F},$ which
appears in Schr\"odinger-Chern-Simons vortex dynamics (\ref{msdyn}),
 is simply proportional to this K\"ahler form, namely 
${\cal F} = -2\omega.$ 

Note that the above formulae for the geometrical objects on the
moduli space are rather implicit, since they involve the 
expansion coefficients $b_r,$ and it is not possible to determine 
these explicitly in terms of the vortex positions $Z_r.$
Furthermore, the reduced potential $V_{\mbox{red}}$, which is the
other ingredient required in the vortex equations of motion (\ref{msdyn}),
is also not available explicitly. It is given by
\be
\label{Vred}
V_{\mbox{red}} = N \pi + \frac{\lambda - 1}{8} \int (1 - |\phi|^2)^2 d^2 x, 
\ee
where the field $\phi$ in the above equation is the solution of 
the Bogomolny equations with zeros at $Z_r$.  
 
Despite the fact that the quantities which appear in the moduli 
space approximation are not known explicitly it is still possible to
derive some interesting results. As an example, it is possible to
show that the naive centre of the $N$-vortex system
\be
Z=\frac{1}{N}\sum_{r=1}^N Z_r
\ee
is a constant of the motion.

The qualitative motion of two vortices can also be understood.
As we have just mentioned, the centre $Z$ is a constant of motion,
which we may choose to be the origin. The vortex positions can then 
be written as $Z_1=-Z_2=\rho e^{i\theta},$ and it follows that
$b_1=-b_2=b(\rho)e^{i\theta}.$ Furthermore, the reduced potential
$V_{\mbox{red}}$ depends only on $\rho.$ In this case the equations of motion
(\ref{msdyn}) become
\begin{eqnarray}
\label{rhodot}
\dot{\rho} &=& 0, \\
\label{thetadot}
2 \pi  \frac{d}{d \rho} 
\left(\rho^2 + 2 \rho~b(\rho) \right) \dot{\theta}
&=& - \frac{\ dV_{\mbox{red}}}{d \rho}.
\end{eqnarray}
These equations imply that
both the vortex separation $2\rho$ and the angular velocity
$\dot{\theta}$ are constant. So, the vortices circle around each other at
constant speed and separation.
If $\lambda<1$ then $dV_{\mbox{red}}/{d \rho}$ is positive, so the
two vortices circle each other clockwise, whereas for $\lambda>1$
the motion is anticlockwise.

An expression for the period $T$ of this circular motion is given by
\begin{equation}
T = \left| \frac{4 \pi^2 \frac{d}{d \rho}
\left(\rho^2 + 2 \rho b(\rho) \right)}{{dV_{\mbox{red}}}/{d \rho}}
\right|.
\label{period}
\end{equation}
Note that the combination $|1-\lambda|T$ is a function only of $\rho,$ 
that is, it is independent of $\lambda,$ so the $\lambda$ dependence
of the period is known. If $\rho$ is large then the speed is 
exponentially small. For small $\rho$ then both the numerator
and the denominator in (\ref{period}) are $O(\rho^3),$ so the period
approaches a finite non-zero limit as $\rho\rightarrow 0.$

\section{Two Vortex Dynamics}\news
In the remainder of this paper we compare the detailed predictions of 
the moduli space approximation with the results of numerical simulations
of the full field theory dynamics. We begin in this section with a 
study of the motion of two vortices.

There are no explicit expressions for the functions $b(\rho)$
and $V_{\mbox{red}}(\rho)$ which appear in the 2-vortex period formula
(\ref{period}), but they can be computed numerically following the 
approaches described in \cite{Sam} and \cite{Shah}. This involves
numerically solving equation (\ref{heqn}) to compute $h,$ for which
we use a gradient flow algorithm with a change of variables that move
the delta function source terms to infinity. The solution $h$ is calculated
for a range of values for the vortex separation $s=2\rho,$ and this is used
both to read-off the function $b(\rho)$ using the definition (\ref{hexpan}),
and to calculate the reduced potential (\ref{Vred}) by performing a
numerical integration over the plane, using the simple relation 
(\ref{htophi}) between $h$ and $|\phi|^2.$ 

\begin{figure}
\begin{center}
\leavevmode \vskip -3.5cm \epsfxsize=15cm\epsffile{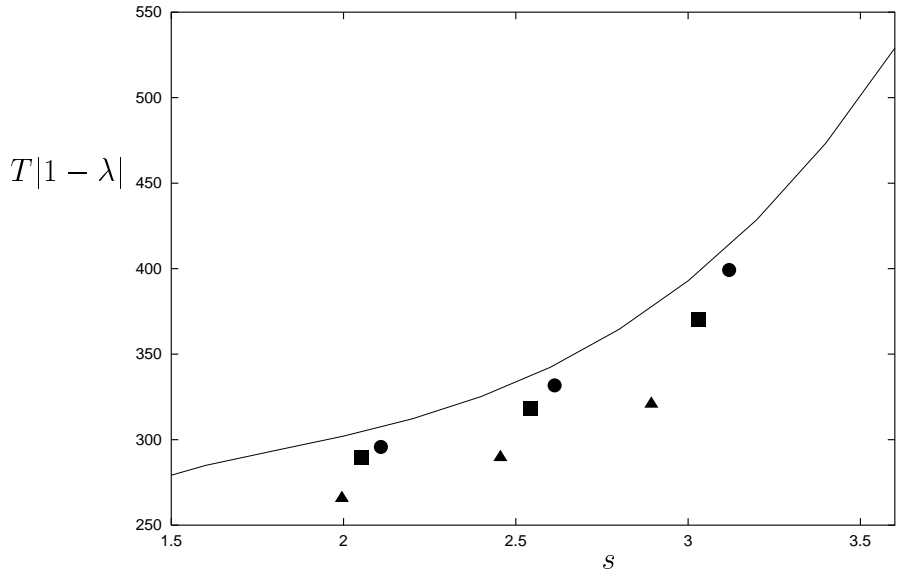} \vskip -10cm
\caption{The scaled period $T|1-\lambda|$ as a function of the vortex
separation $s,$ as predicted by the moduli space approximation. The data
points show the scaled period as calculated from full field theory simulations
with couplings 
$\lambda=0.9$ (circles), $\lambda=0.75$ (squares), $\lambda=0.5$ 
(triangles). }
\label{fig-period}
\end{center}
\end{figure}

As mentioned in the previous section, the combination 
$|1-\lambda|T,$ which we refer to as the scaled period,
is independent of $\lambda.$ It is this quantity that is plotted 
in Fig.~\ref{fig-period}, as a function of the separation $s=2\rho,$
for $1.5\le s\le 3.6.$ This graph displays the expected feature
 that the period of rotation increases with the vortex separation.

Recall that the period tends to a finite
non-zero value as the separation tends to zero, and it would be useful
if our numerical computation could be extended to the range $0\le s\le 1.5,$
so that this value could be computed. However, recall that the period
is calculated as the ratio of two quantities which are both $O(\rho^3)$
as $\rho$ tends to zero. Furthermore, both of these quantities 
involve differentiating terms which we compute numerically. Thus, although
the numerically computed functions $b(\rho)$ and $V_{\mbox{red}}(\rho)$
are reasonably accurate, calculating the period for small separations is much
more sensitive to numerical errors, since it involves the ratio of
two terms whose quartic approaches to zero need to be found. This is why
our calculation for the period is only presented for $\rho\ge 0.75,$ since
for smaller values we are not able to achieve the accuracy required.
To improve the accuracy the solution $h$ of the partial differential equation
(\ref{heqn}) would need to be solved on larger grids than we currently
use, but this is computationally rather expensive.
  
An alternative approach would be to linearize around the axially
symmetric two vortex solution in order to obtain the period of rotation for 
minimal separation. Since the two vortex solution is not known analytically,
this is a nontrivial numerical problem, but it is an interesting project  
for further study.

We now turn to numerical simulations of the full field theory dynamics.
The evolution equations (\ref{Schr}) and (\ref{Amp}) are solved using
a second order finite difference scheme for the spatial derivatives
with lattice spacing $dx=0.1$ on a $400\times 400$ grid. The time
evolution uses a fourth order Runge-Kutta algorithm with a time step
$dt=0.005.$ Note that the time step is much smaller than the lattice
spacing, as expected for the numerical solution of a partial differential
equation which is first order in time derivatives and second order
in spatial derivatives.

\begin{figure}
\begin{center}
\leavevmode \vskip -3.5cm \epsfxsize=15cm\epsffile{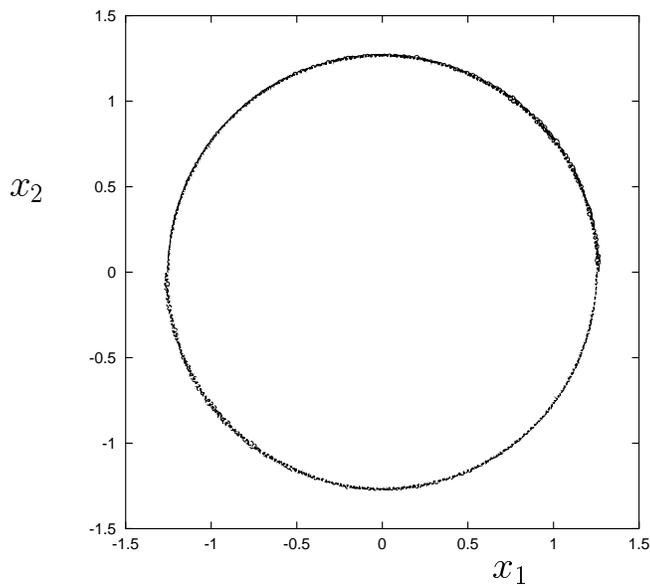} \vskip -10cm
\caption{The vortex positions for two vortices with initial separation
$s=2.5$ and coupling $\lambda=0.9.$ }
\label{fig-0.9}
\end{center}
\end{figure} 

Recall that the Gauss law equation (\ref{Gauss}) is a constraint
on the initial data. This means that it is a non-trivial
task to construct initial conditions for simulations; in particular
the usual method of a simple product ansatz for well separated vortices 
can not be used since this will not satisfy the constraint.
One of the main motivations for our work is to compare field theory
dynamics with the moduli space approximation, and this suggests 
a natural solution to this problem. As initial conditions we use solutions
of the Bogomolny equations (\ref{Bog1}) and (\ref{Bog2}), which are static
solutions only at critical coupling, since these automatically satisfy
the Gauss law constraint (\ref{Gauss}). In summary, to obtain initial
conditions for our simulations we fix the vortex positions, then solve
equation (\ref{heqn}) to obtain the field $h$ with these vortex positions,
and finally reconstruct the initial condition fields $\phi$ and $a_i$ 
from $h,$ setting the initial value of $a_0$ to zero. 

The results of a two vortex simulation with $\lambda=0.9$ and
initial positions $(x_1,x_2)=(\pm 1.25,0)$, so that the  
initial separation is $s=2.5,$ are presented in Fig.~\ref{fig-0.9}.
The vortex positions are defined as the locations of the zeros of the
Higgs field, and it is these that are shown in Fig.~\ref{fig-0.9} for
times $0\le t\le 1646.$ The two vortices circle around each other clockwise
at approximately constant speed and separation. The results are displayed
for a half-period giving $T/2=1646,$ which corresponds to more than 
$300,000$ timesteps with $dt=0.005.$ The motion is very slow because
the coupling $\lambda=0.9$ is near to the critical value, so the forces
are weak. 

Clearly this simulation confirms 
the qualitative prediction of the moduli space approximation,
that two vortices circle around each other.
To compare the quantitative prediction we plot 
the period of this motion against the separation as a circle
data point on Fig.~\ref{fig-period}. The separation is slightly larger
than $2.5$ since we calculate it not from the initial value, but as
a value averaged over the half-period, and the vortices very slightly drift
apart. The additional circles in  Fig.~\ref{fig-period} represent the
results of similar computations to calculate the period for different
values of the initial separation. The circles in  Fig.~\ref{fig-period}
are reasonably close to the solid curve prediction of the moduli space
approximation, indicating that at this value of the coupling it provides
good results.

\begin{figure}
\begin{center}
\leavevmode \vskip -3.5cm \epsfxsize=18cm\epsffile{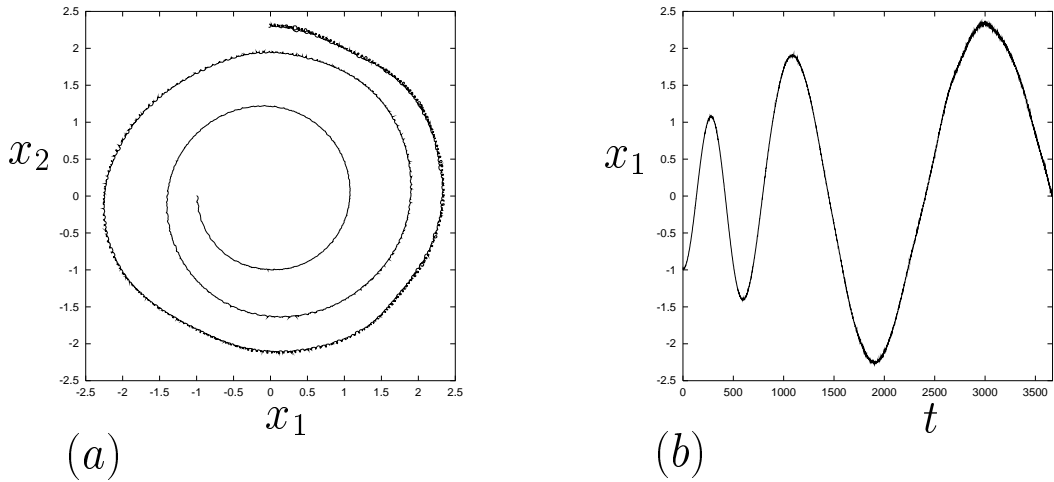} \vskip -15cm
\caption{The position of one of the vortices, for times
$0\le t \le 3670,$ for a two vortex simulation
 with initial separation $s=2$ and coupling $\lambda=1.5$. (a) The position in
the $(x_1,x_2)$ plane. (b) The $x_1$ component of the position as
a function of time.}
\label{fig-1.5}
\end{center}
\end{figure}

In Fig.~\ref{fig-1.5} we present the results of a simulation with 
initial separation $s=2$ and coupling $\lambda=1.5.$ For clarity we
display the position of only one of the vortices; the position of the
other vortex being obtained by symmetry as $(x_1,x_2)\mapsto(-x_1,-x_2).$
Fig.~\ref{fig-1.5}(a) shows the position in the $(x_1,x_2)$ plane
for times $0\le t \le 3670,$ and Fig.~\ref{fig-1.5}(b) displays
the $x_1$ component of the position as
a function of time.

From these figures it can be seen that this time the vortices circle 
around each other anticlockwise, as
predicted by the moduli space approximation when $\lambda>1.$
However, the separation is far from being constant and increases 
substantially with time, producing an outward spiral motion of the vortices.
The outward motion of the vortices is the same order of magnitude as 
the circular motion, indicating that there are significant forces
acting on the vortices that are neglected in the moduli space approximation.
Fig.~\ref{fig-1.5}(b) displays the $x_1$ component of the position as
a function of time, verifying that the time taken for the vortices to
circle each other increases as they move further apart.

\begin{figure}
\begin{center}
\leavevmode \vskip -3.5cm \epsfxsize=18cm\epsffile{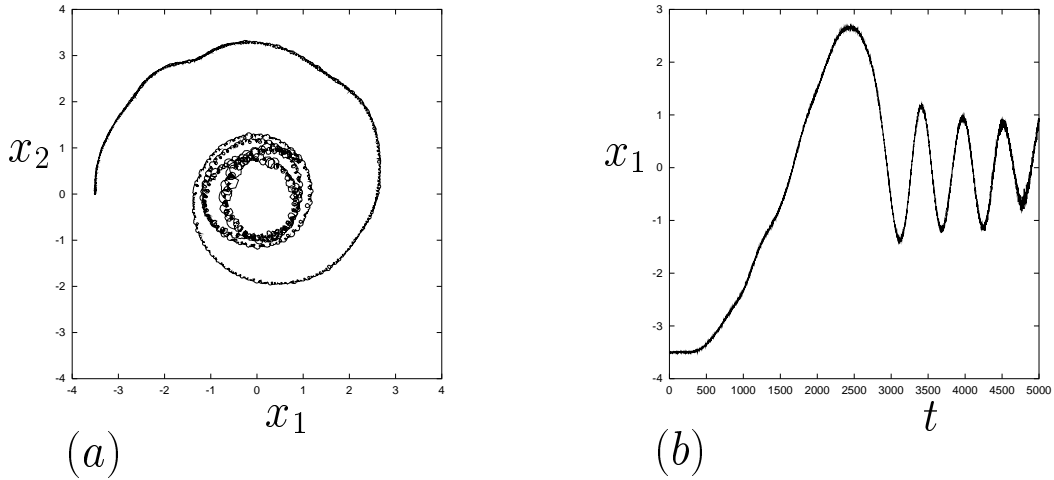} \vskip -15cm
\caption{The position of one of the vortices, for times
$0\le t \le 5000,$ for a two vortex simulation
 with initial separation $s=7$ and coupling $\lambda=0.5$. (a) The position in
the $(x_1,x_2)$ plane. (b) The $x_1$ component of the position as
a function of time.}
\label{fig-0.5}
\end{center}
\end{figure}

A natural interpretation of the outward spiral motion is that the 
vortices radiate as they circulate. Recall from Section \ref{sec-model} that
for $\lambda>1$ the energy of two vortices decreases with increasing 
separation, so an energy loss due to radiation would result in the 
vortices moving away from each other. If this is indeed the correct
explanation of the outward spiral motion, then it predicts that for
couplings not too close to the critical value and satisfying $\lambda<1$ then
the vortices should circle around each other with an inward spiral motion.
This is because for $\lambda<1$ the results from Section \ref{sec-model}
show that the energy of two vortices decreases as the separation
decreases, so this time an energy loss due to radiation would result in the 
vortices moving towards each other. 
In order to test this prediction we
perform a simulation with initial separation $s=7$ and coupling
$\lambda=0.5.$ The results are displayed in Fig.~\ref{fig-0.5} using
the same format as in Fig.~\ref{fig-1.5}.

The vortices circle each other clockwise and the inward spiral motion is 
clearly visible in Fig.~\ref{fig-0.5}. This is consistent with the
above expectation based on an energy loss due to radiation.
 The $x_1$ component of the position, plotted in Fig.~\ref{fig-0.5}(b), 
suggests that the energy loss will eventually lead to the final configuration
being a static $N=2$ axial vortex, though the approach to this static
solution is very slow. The simulation presented already involves one
million time steps of the evolution algorithm, so it is computationally
too expensive to attempt to follow the evolution for any substantially
longer time. There is always the possibility that the configuration
could stabilize to a rotation at a very small non-zero separation, but
this seems unlikely.  

It is evident from Fig.~\ref{fig-0.5}(a) that for later times there
is a significant wobbling of the vortex motion. This wobbling component
eventually appears in all our simulations. It can be seen in
 Fig.~\ref{fig-0.9} and Fig.~\ref{fig-1.5}, though it is much less
pronounced in these cases. We do not know whether this is a 
purely numerical artifact or has some physical significance. It certainly
signals the onset of a numerical instability in our evolution algorithm
since our codes are unable to continue the evolution for very long after
a pronounced wobbling emerges. Monitoring the constraint (\ref{Gauss})
we find that it is satisfied to within a reasonable numerical accuracy
until shortly before the evolution algorithm fails, but we do not know
whether a drift away from this constraint is a cause or symptom of
the instability. There is an initial release of radiation at the start
of all our simulations, since the initial conditions are created for
vortices with critical coupling whereas the simulations require a non-critical
value. There is some evidence, from changing grid sizes, 
that the interaction of this radiation with the boundary of the grid may 
be a factor in limiting our simulation times. We use Neumann boundary
conditions, which results in some radiation reflecting from the
boundary. It would be better if absorbing boundary conditions could be
used, but this is not easy to implement since radiation in the
Schr\"odinger-Chern-Simons equations is non-standard.
Linearizing the evolution equations shows that the plane wave
solutions are elliptically polarized. See the appendix for further details.

In summary, we have seen that for two vortices the moduli space prediction
of circular motion of the vortices around each other is only a good
approximation very close to critical coupling.    
For modest deviations from the critically coupled regime there is a 
substantial spiral motion, which is an inward spiral for Type I vortices
 and an 
outward spiral for Type II. Very close to critical coupling we found
that the moduli space approximation for the period of the circular
motion was quite accurate. Further away from critical coupling the 
motion is not even approximately periodic, but an indication of the
time scale of the rotation can be computed by calculating the \lq period\rq\  
as twice the time taken for the vortices to make a rotation through
$180^\circ.$ We include values of these scaled periods in 
Fig.~\ref{fig-period} for various separations and couplings $\lambda=0.5$
(triangles) and $\lambda=0.75$ (squares). Comparing these values
with the near critical value $\lambda=0.9$ (circles) suggests
that the moduli space prediction for the period is indeed approached as
the coupling tends to the critical value $\lambda=1.$  

\section{Rotating Polygons}\news
Symmetry of the moduli space dynamics (\ref{msdyn}) implies that
if there are $N$ vortices at the vertices of a regular $N$-gon
then this polygon of vortices will rigidly rotate about its
centre with constant size and speed \cite{Ma10}.

The linear stability of such a rotating polygon has been studied within
the moduli space approximation \cite{SpR}. This situation is a 
generalization of the two vortex system, so it is not surprising
that neither the K\"ahler form nor the reduced potential are explicitly
known. The approach of Ref.\cite{SpR} is to consider the limit in which
the size of the polygon is large, so that the vortices are well separated.
For well separated vortices there is an asymptotic formula for the
K\"ahler form \cite{MaSp} and the reduced potential can be approximated
by the Ginzburg-Landau energy, for which there is an asymptotic formula
in terms of the sum over the 
asymptotic two-vortex interaction energies \cite{Sp1}.
With all these approximations, the result \cite{SpR} is that the
$N$-gon is stable if and only if $N<6.$ 

Given the field theory simulations described in the previous Section, 
together with the energy calculations of Section \ref{sec-model},
we expect that symmetric polygons will spiral in for $\lambda<1$
and out for $\lambda>1,$ so in that sense they are all unstable.
However, the stability results referred to above still apply if
we ask about the stability of the symmetry of the configuration, that is,
we shall refer to the $C_N$ symmetric polygon as being stable
if the $C_N$ symmetry is stable, irrespective of a size instability.
  
In this Section we present the
results of full field theory simulations on rotating polygons, to
compare with the stability predictions of the moduli space approximation.
These predictions are most likely to hold close to critical coupling
and for polygons of a large size, so that the vortices are well
separated. Unfortunately, both these regimes are difficult to simulate
numerically since the forces are very weak. This makes the vortex
motion very slow and requires an unacceptably long simulation time.
Nonetheless, we find that there is a substantial agreement between
the moduli space predictions and the results of our simulations, though
there is a disagreement for a pentagonal arrangement.

The simulations we present are with a coupling constant $\lambda=0.5,$
though we also obtained similar results with other values. 
There is no need to explicitly break the cyclic symmetry 
of the vortex positions,
since the numerical discretization, and in particular the 
boundary of the grid, are sufficient to allow the symmetry to break if
there is an unstable mode. 

\begin{figure}
\begin{center}
\leavevmode \vskip -3.5cm \epsfxsize=15cm\epsffile{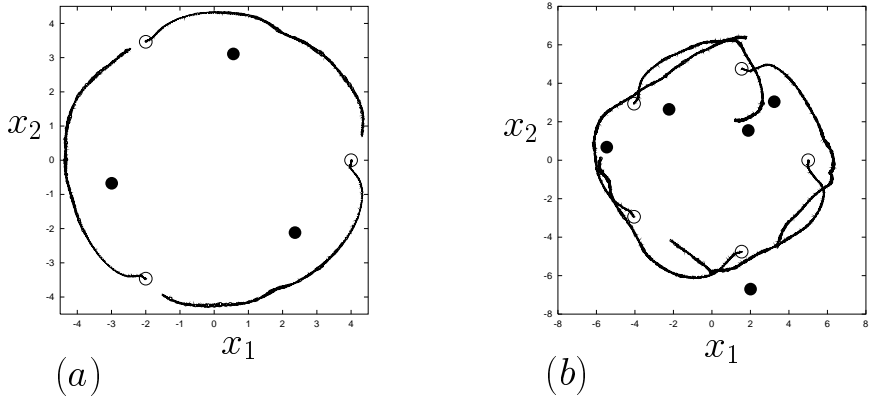} \vskip -13cm
\caption{Vortex positions for polygonal initial conditions. The white circles
indicate the initial vortex positions and the black circles indicate
the final vortex positions at the end of the simulation. The lines show
the vortex positions in the early stages of the simulation.
(a) A triangular arrangement with vortices initially on a circle of
radius 4. Lines are drawn for times $t\le 2000$ and the black circles
are the positions at the time $t=5000.$
(b) A pentagonal arrangement with vortices initially on a circle of
radius 5. Lines are drawn for times $t\le 2500$ and the black circles
are the positions at the time $t=3200.$
}
\label{fig-N35}
\end{center}
\end{figure}

In Fig.~\ref{fig-N35}(a) we display the vortex positions for an
initial triangular arrangement with three vortices on a circle of
radius 4. The white circles denote the initial positions and the
lines indicate the resulting motion for times $t\le 2000.$ For clarity
we do not display the vortex positions as lines for later times, but
do show the positions (black circles) at the end of the simulation
 at time $t=5000.$ At the end of the simulation the vortices have made
just over one full rotation, and still preserve the triangular symmetry.
Note that at $t=2000$ the vortices have made less than one third of
a full rotation, but at $t=5000$ they have made just over a full rotation.
This is because the vortices radiate and their separation decreases, with
a corresponding decrease in the rotation period. This simulation is
consistent with the asymptotic moduli space prediction that a triangular
arrangement is stable.

An arrangement of four vortices on the vertices of a square produces
a similar result, with the vortices moving on an inward spiral but
preserving the $C_4$ symmetry.

\begin{figure}
\begin{center}
\leavevmode \vskip -1cm \epsfxsize=15cm\epsffile{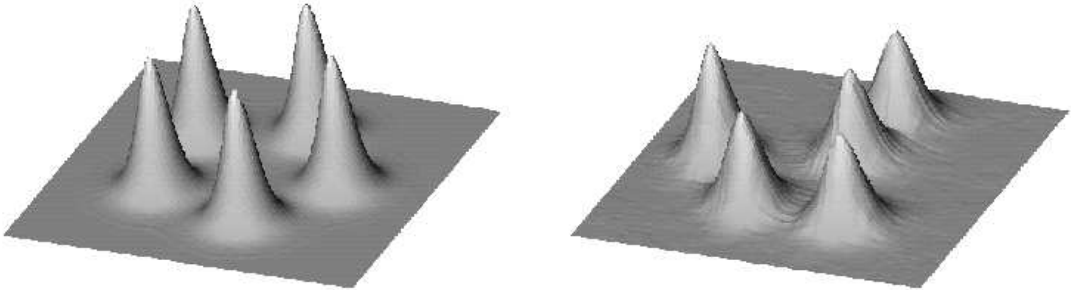} \vskip -0cm
\caption{Energy density plots for an initial pentagonal arrangement
of 5 vortices. The plots are at times $t=0$ and $t=2500.$}
\label{fig-c5}
\end{center}
\vskip -1cm
\end{figure}

\begin{figure}
\begin{center}
\leavevmode \vskip -1cm \epsfxsize=15cm\epsffile{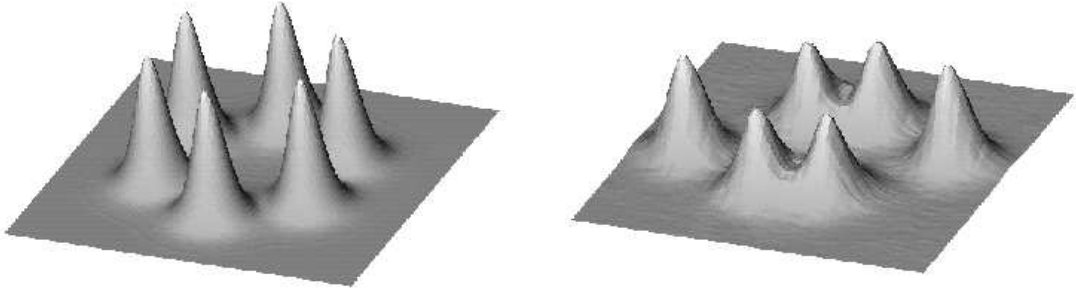} \vskip -0cm
\caption{Energy density plots for an initial hexagonal arrangement
of 6 vortices. The plots are at times $t=0$ and $t=2000.$}
\label{fig-c6}
\end{center}
\vskip -0.5cm
\end{figure}

The results of a simulation with five vortices on the vertices of a regular
pentagon are shown in Fig.~\ref{fig-N35}(b). The initial vortex positions
(white circles) lie on a circle of radius 5, and the lines track the
motion for $t\le 2500.$ It can already be seen from these vortex trajectories
that the cyclic $C_5$ symmetry gets broken, with one of the vortices
already moving closer to the origin than the other four. The black circles
denote the vortex positions at time $t=3200,$ and the pentagonal symmetry
has obviously been destroyed, with one of the vortices being much
further from the origin than the other four. 
The broken symmetry is most easily visualized through energy density
plots. Fig.~\ref{fig-c5} shows energy density plots at times 
$t=0$ and $t=2500,$ clearly displaying the broken $C_5$ symmetry.
Note that the vortices are more localized in the initial conditions
than at later times. This is because the initial conditions are formed
from critically coupled vortices, but these quickly relax to vortices
with a localization appropriate for the chosen coupling $\lambda=0.5.$ 
The numerical grid extends far beyond the inner portion plotted
in these figures, but for clarity we do not display the entire grid.
 
Further evolution of this
configuration suggests that the vortices are tending towards the formation
of a pair of axial $N=2$ vortices, plus a single $N=1$ vortex, though
our simulations can not be run long enough to confirm the final state.
A reasonable expectation is that the vortices combine, 
at first in pairs, and eventually radiation loss will lead to a final
configuration of an axial $5$-vortex.

The instability of the pentagonal arrangement disagrees with the
asymptotic moduli space prediction and therefore raises a number of
interesting questions as to the source of the disagreement. One possibility
is that the asymptotic approximation is correct but that the simulation 
is not in the asymptotic regime. This would suggest that there may be 
a critical size for the pentagonal arrangement, with a stability only
if the pentagon is above the critical size. 
This could be investigated by numerically computing the appropriate
K\"ahler form and reduced potential, and then numerically solving an
eigenvalue problem to determine if there are any negative modes.
This is an interesting project for future investigation.

Another possibility is that
there is a stability region in coupling space, so that the pentagon
is stable only if $\lambda$ is sufficiently close to critical coupling.
It might be possible to investigate this within a moduli
space approach by including the deformations of the vortex fields, 
such as the tiny electric field, which arise away from critical coupling
and are not captured by restricting to the critically coupled moduli space.
For $\lambda>1$ an appropriate moduli space could be constructed from the
 unstable manifold of the axial $N=5$ vortex, in a similar way that
has been used \cite{LaSc} to study Schr\"odinger dynamics of two ungauged
Ginzburg-Landau vortices.  

A final possibility is that the instability is due to a phenomenon
that can not be captured by any standard moduli space approach, and the
interaction between vortices and radiation plays a crucial role.

The asymptotic moduli space approach predicts that 6 vortices 
on the vertices of a regular hexagon will be unstable.
The results of such an initial arrangement, with the vortices on
a circle of radius 5, are presented in Fig.~\ref{fig-c6}.
This figure displays energy density plots at times $t=0$ and $t=2000.$
Clearly the cyclic $C_6$ symmetry is broken by this time, so this
result is in agreement with the instability prediction. 
Note from the second plot in Fig.~\ref{fig-c6} that the $C_6$
symmetry is broken to a $C_2$ symmetry at this stage
in the evolution, and again there appears to be a tendency for vortices
to form pairs. 

In summary, we find that the stability properties of regular polygonal
arrangements of vortices is in
reasonable agreement with the moduli space prediction, in that
we find that an $N$-gon is stable only for small enough values
 $N<N_c.$ However, we conclude that the critical value is $N_c=5$
rather than $N_c=6,$ and have suggested some approaches
that could be used to investigate this discrepancy further. 
 
\section{Hyperbolic Vortices}\news
The equations for static critically coupled 
Ginzburg-Landau vortices on the hyperbolic
plane of curvature $-\frac{1}{2}$ are integrable, and this allowed
Strachan \cite{Str} to calculate a general formula for the K\"ahler
form on the moduli space $M_N.$ 
In this respect hyperbolic vortices are therefore 
simpler than Euclidean vortices. In this Section we exploit this fact to study
the moduli space approximation to Schr\"odinger-Chern-Simons 
vortex dynamics on hyperbolic space.

Consider a two-dimensional Riemannian manifold $X,$ with metric 
\begin{equation}
ds^2 = \Omega(x_1,x_2) (dx_1^2 + dx_2^2).
\end{equation}
Ginzburg-Landau vortices on $X$ are minima of the energy
\begin{equation}
\label{V}
V = \frac{1}{2} \int_X \left(
\Omega^{-1} B^2 +\overline{D_1 \phi} D_1 \phi + \overline{ D_2 \phi} D_2 \phi 
+ \frac{\lambda \Omega}{4} (1- \bar{\phi} \phi)^2 \right) d^2 x.
\end{equation}
At the critical coupling $\lambda=1$ the second order equations
can again be reduced to first order Bogomolny equations, as in
the Euclidean case. Introducing the function $h,$ as in (\ref{htophi}),
then the generalization of the Euclidean equation (\ref{heqn})
becomes
\begin{equation}
\nabla^2 h + \Omega - \Omega {\rm e}^h = 4 \pi \sum\limits_{r=1}^N 
\delta(x - X_r), 
\end{equation}
where $\nabla^2$ is the standard flat space Laplacian and
$X_r\in X$ are the vortex positions.

In the Poincar\'e disc model, the metric of the hyperbolic plane
with curvature $-\frac{1}{2}$ is
\begin{equation}
ds^2 = \frac{8}{(1-|z|^2)^2} dz d{\bar z} = \Omega~ dz d{\bar z},
\end{equation}
where $|z| < 1$.
Setting $h = 2 g + 2\log \frac{1}{2}(1-|z|^2)$ the
equation for $h$ becomes Liouville's equation with sources
\begin{equation}
\nabla^2 g - {\rm e}^{2g} = 2\pi\sum\limits_{r=1}^N 
\delta(z - Z_r), 
\end{equation}
which can be solved exactly. The solution is
\begin{equation}
g = - \log \frac{1}{2}(1-|f|^2) + 
\frac{1}{2} \log \left|\frac{df}{dz}\right|^2,
\end{equation}
where $f(z)$ is an arbitrary, complex analytic function.
 With a simple
choice of phase the scalar field is given by
\begin{equation}
\phi = \frac{1 - |z|^2}{1-|f|^2}\frac{df}{dz}.
\label{phisoln}
\end{equation}
Note that $\phi$ vanishes at the zeros of $\frac{df}{dz}$.
To ensure that $\phi$ has $N$ zeros, is
nonsingular inside the disc $|z| <1$ and is zero on the boundary
$|z|=1,$ then the function $f(z)$ needs to have the Blaschke product
form 
\begin{equation}
f(z) = \prod\limits_{i=1}^{N+1} \left( \frac{z - a_i}{1 - {\bar a}_i z}
\right),
\end{equation}
where $|a_i| < 1$. 

Let us now specialize to the case of $N$ vortices on an $N$-gon,
including the degenerate case $N=2.$ 
Such a configuration is described by the holomorphic
function 
\begin{equation}
\label{fmn}
f(z) = \frac{z\left(z^N-a^N\right)}{1-{\bar a}^N z^N}.
\end{equation}
The positions of the vortices are given by the zeros of the derivative 
${df}/{dz}$, and these are the regular $N$-gon vertices given by 
\begin{equation}
z = \alpha~{\rm e}^{2 \pi i k/N} 
\end{equation}
for $k=1,\dots, N$. The parameter $a$ can be expressed in terms of 
$\alpha$ via $a = \alpha \gamma$ where
\begin{equation}
\gamma^N = \frac{ 
\sqrt{\left(1-|\alpha|^{2N} \right)^2 + 4 N^2 |\alpha|^{2N}}
 -1 - |\alpha|^{2N}}
{2 |\alpha|^{2N} \left(N-1 \right)}.
\end{equation}
Using the results of \cite{Str} the K\"ahler 
metric on this two-dimensional submanifold of $M_N$ can be calculated 
explicitly 
and is given by
\begin{equation}
\label{m1}
ds^2 = \frac{4 \pi N^3 |\alpha|^{2N-2} d\alpha~d{\bar \alpha}}
{\left(1-|\alpha|^{2N} \right)^2}
\left( 1 + 
\frac{2 N \left(1 + |\alpha|^{2N} \right)}
{\sqrt{\left(1-|\alpha|^{2N} \right)^2 + 4 N^2 |\alpha|^{2N}}}
\right).
\end{equation}
Note that for $N=2$ this metric becomes the one derived by
Strachan \cite{Str}, which he used to study the relativistic
dynamics of two hyperbolic vortices.

Given the metric (\ref{m1}) the  K\"ahler form $\omega$ can simply be read off.
Let $\alpha = r {\rm e}^{i \theta},$ then 
\begin{equation}
\label{k}
\omega = f(r) dr \wedge d\theta,
\end{equation} 
where 
\begin{equation}
f(r) = \frac{2 \pi N^3 r^{2N-1}}
{\left(1-r^{2N} \right)^2}
\left( 1 + 
\frac{2 N \left(1 + r^{2N} \right)}
{\sqrt{\left(1-r^{2N} \right)^2 + 4 N^2 r^{2N}}}
\right).
\end{equation}

Now that we have the K\"ahler form, all that we require 
to study  Schr\"odinger-Chern-Simons vortex dynamics
is the reduced potential. This is given by
the expression
\begin{equation}
\label{Vred2}
V_{\mbox{red}} = N\pi +\frac{\lambda -1}{8} \int_X (1 - |\phi|^2)^2~
\Omega ~ d^2 x,  
\end{equation}
where the explicit solution for $\phi$ is known from above using 
(\ref{phisoln}) and
(\ref{fmn}).
It can be shown that this reduced potential depends only on $|\alpha|.$
Unfortunately this integral over the hyperbolic plane can not be 
performed explicitly (except for
the special value $\alpha=0$) even in the case $N=2.$ The angular part of 
the integral can be calculated explicitly but the radial integration
must then be computed numerically. 

With the K\"ahler form and reduced potential given above the 
moduli space dynamics reduces to the equations of motion
${\dot r} = 0$ and 
\begin{equation}
2 f(r) {\dot \theta} = -\frac{dV_{\mbox{red}}}{dr}.
\end{equation}
Thus, as in the Euclidean case, the configuration rigidly rotates 
with no change in size and
 with constant angular velocity ${\dot \theta}.$ The period is
\begin{equation}
T = \left|\frac{4 \pi f(r)}{d V_{\mbox{red}}/dr}\right|.
\end{equation}

In order to compare the period calculation in hyperbolic space to
 the earlier results in Euclidean space, we need to calculate the
proper distance. 
From the Poincar\'e disc coordinate $z=|z|e^{i\chi}$
we make the transformation 
$\rho  = 2 \sqrt{2} \tanh^{-1}(|z|)$ so that the hyperbolic metric becomes
\begin{equation}
ds^2 = d\rho^2 + 2 \sinh^2 \left(\frac{\rho}{\sqrt{2}}\right) d\chi^2.
\end{equation}
Thus $\rho$ should be equated with the radial distance in Euclidean space.

\begin{figure}
\begin{center}
\leavevmode \vskip -3cm 
\epsfxsize=15cm\epsffile{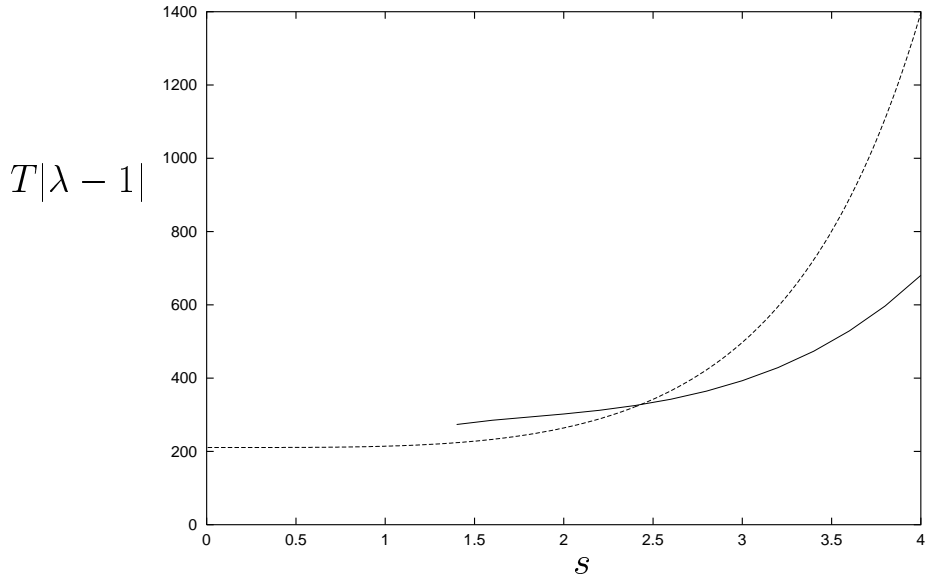} 
\vskip -10.5cm
\caption{The scaled period $T|\lambda -1|$ against separation $s$ for
two vortices in Euclidean space (solid curve) and hyperbolic
space (dashed curve).}
\label{fig-hyp}
\end{center}
\end{figure}

In Fig.~\ref{fig-hyp} we plot (dashed curve) the scaled period
 $T|\lambda -1|$ as a 
function of the separation $s=2\rho$ for the two-vortex case ie. $N=2.$ 
For comparison we
also reproduce the earlier Euclidean result (solid curve). 
The Euclidean and hyperbolic results are qualitatively similar.
Both periods are monotonically increasing functions of the separation
and tend to finite non-zero limits as the separation tends to zero.
The simplifications of hyperbolic space allow the period to be 
calculated accurately all the way down to zero separation, and this value
appears similar to the one extrapolated from the Euclidean data. Note that
it is not expected that the Euclidean and hyperbolic periods agree
at zero separation since the vortices are extended objects, so even when
both vortices are located at the origin they still feel the curvature
of hyperbolic space.  

The period for an $N$-gon arrangement of hyperbolic vortices can
also be calculated using the above methods and the results
are qualitatively similar to the 2-vortex case. The period at zero
separation is not very sensitive to the value of $N$ but does slightly
decrease as $N$ increases.

The similarity between the moduli space results for Euclidean and
hyperbolic space suggests that it would be interesting to investigate
the stability of polygonal arrangements of hyperbolic vortices.
Although this is still not an easy exercise it is substantially
simpler than the Euclidean case, since the exact static vortex solutions
are available at critical coupling.

\section{Conclusions}\news
In this paper we have studied the dynamics of vortices in Manton's
Schr\"odinger-Chern-Simons model, and compared the results of
full field simulations with predictions from the moduli space
approximation. We found that there is a good agreement for
couplings extremely close to the critical coupling, but away from this
value there are significant qualitative differences, which we attribute
to radiation effects which are not captured by the moduli space
approach. This is of physical relevance since real superconductors
are generally not close to the Type I/Type II transition.

A novel suggestion might be to try and modify the non-dissipative moduli space
dynamics to a new dissipative dynamics on the moduli space that would
capture the energy loss effects. The Schr\"odinger-Chern-Simons flow is
orthogonal, in a rigourous mathematical sense, to the gradient flow 
of the Ginzburg-Landau energy, so a flow which contains both components
should certainly reproduce the qualitative features that we have found.
However, it is not clear how to derive such a flow from the field theory,
but it would seem a worthwhile avenue for further investigation.

In the Schr\"odinger dynamics of two ungauged
Ginzburg-Landau vortices it has been shown analytically 
that two vortices radiate while rotating around each other and
the asymptotic rate at which they radiate has been derived \cite{OvSi}. 
In the ungauged system the coupling constant plays no role. Motivated
by the numerical results we have presented here, it would be interesting
if the methods used in \cite{OvSi} could be applied to the 
Schr\"odinger-Chern-Simons model and similar results derived, including
the coupling constant dependence. This is another interesting topic
for future work, though the inclusion of gauge fields, and in particular the
requirement of working with gauge independent quantities, appears to
make the problem significantly more complicated than in the ungauged
situation.

Finally, we have demonstrated that, at least within the
moduli space approximation, the qualitative features of Euclidean vortices
appear to be shared by hyperbolic vortices. This opens up the possibility
of future studies on the stability properties of symmetric arrangements
of hyperbolic vortices, for arbitrary separations, exploiting the fact
that hyperbolic vortices are simpler to study due to the existence
of exact static solutions at critical coupling. 

\renewcommand{\theequation}{A.\arabic{equation}}
\section*{Appendix: The linearized field equations}\news

In this appendix, we discuss the linearized field equations and their
plane wave solutions. It is convenient to work in a gauge such that
$\phi$ is real. Setting $\phi = 1 + \sigma$, we can linearize around
the vacuum $\phi = 1$ and $a_\mu = 0$, by expanding the field
equations (\ref{Schr}), (\ref{Amp}) and (\ref{Gauss}) in $\sigma$ and
$a_\mu$ to linear order. First, we eliminate $a_0$ using the
linearized real part of (\ref{Schr}) 
\begin{equation}
2 a_0 = -\nabla^2 \sigma + \lambda \sigma,
\end{equation} 
where $\nabla^2=\partial_1^2+\partial_2^2.$
Then we replace $\sigma$ using the linearized Gauss equation
\begin{equation}
\sigma=-B.
\end{equation}
Finally, from (\ref{Amp}) we obtain
\begin{equation}
\label{waveequation}
\left(
\begin{array}{c}
2 {\dot a_1} \\
2 {\dot a_2}
\end{array} 
\right)
= 
\left(
\begin{array}{cc}
(1+\lambda-\nabla^2) \partial_1 \partial_2  &
1- (1+\lambda-\nabla^2) \partial_1^2\\
-1+ (1+\lambda-\nabla^2) \partial_2^2 &
-(1+\lambda-\nabla^2) \partial_1 \partial_2 
\end{array}
\right)
\left(
\begin{array}{c}
a_1 \\ a_2
\end{array}
\right).
\end{equation}
Note that the determinant of the symbol of the differential operator
in (\ref{waveequation}) vanishes. In order to find
plane wave solutions we make the ansatz $a_i = A_i \exp(i ({\bf k}
\cdot {\bf x} - \omega t))$. This results in a linear  
homogeneous matrix equation which has
nontrival solutions only if the determinant vanishes, namely,
\begin{equation}
- 4 \omega^2 + k^4 + (1 + \lambda) k^2 + 1 = 0,
\end{equation}
where $k^2 = k_1^2 + k_2^2$.
This gives rise to the nonlinear dispersion relation
\begin{equation}
\omega = \frac{1}{2} \sqrt{k^4 + (1+ \lambda) k^2 + 1},
\end{equation}
which for critical coupling $\lambda = 1$ reduces to
\begin{equation}
\omega = \frac{1}{2} (k^2 + 1). 
\end{equation} 
For general $\lambda$ the corresponding linearized solution is given by 
\begin{equation}
\label{wavevector}
 \left(\begin{array}{c} a_1 \\ a_2 \end{array}\right)=
\alpha \left(
\begin{array}{c}
\cos \left({\bf k} \cdot {\bf x} - \omega t + \delta \right) \\
\frac{k_1 k_2 (k^2 + 1 + \lambda)}{k_1^2(k^2+1+\lambda)+1}
\cos \left({\bf k} \cdot {\bf x} - \omega t + \delta \right)
+ \frac{\sqrt{1+(1+\lambda)k^2+k^4}}{k_1^2(k^2+1+\lambda)+1}
\sin \left({\bf k} \cdot {\bf x} - \omega t + \delta \right)
\end{array}
\right),
\end{equation}
where $\alpha$ and $\delta$ are constants. 
This shows that the plane wave solutions are elliptically polarised,
as can be seen for example from the fact that the modulus of the vector 
(\ref{wavevector}) is not time independent.

\section*{Acknowledgements}
Many thanks to Martin Speight for several helpful discussions 
at the early stages of this project and
to Nick Manton for useful comments.
 This work was supported by the PPARC special programme
grant ``Classical Lattice Field Theory''. SK acknowledges the EPSRC
for a postdoctoral fellowship GR/S29478/01.

\end{document}